\documentstyle[preprint,aps]{revtex}
\tightenlines
\begin{document}
\title{Simple model of 1/f noise}
\author{B. Kaulakys}
\address{Institute of Theoretical Physics and Astronomy, A. Go\v stauto 12, 2600\\
Vilnius, Lithuania }
\date{Received 1 March 1998 }
\maketitle

\begin{abstract}
Simple analytically solvable model of 1/f noise is proposed. The model
consists of one or few particles moving in the closed contour. The drift
period of the particle round the contour fluctuates about some average
value, e.g. due to the external random perturbations of the system's
parameters. The model contains only one relaxation rate, however, the power
spectral density of the current of particles reveals an exact 1/f spectrum
in any desirable wide range of frequency and can be expressed by the Hooge
formula. It is likely that the analysis and the generalizations of the model
can strongly influence on the understanding of the origin of 1/f noise.

\noindent PACS number(s): 05.40.+j, 02.50.-r, 72.70.+m
\end{abstract}


The puzzle of the origin and omnipresence of 1/f noise -- also known as
'flicker' or 'pink' noise -- is one of the oldest unsolved problem of the
contemporary physics. Since the first observation of the flicker noise in
the current of electron tube more than 70 years ago by Johnson \cite
{johnson25}, fluctuations of signals and physical variables exhibiting
behavior characterized by a power spectral density $S\left( f\right) $
diverging at low frequencies like $1/f^\delta $ ($\delta \simeq 1$) have
been discovered in large diversity of uncorrelated systems. We can mention
here processes in condensed matter, traffic flow, quasar emissions, music,
biological, evolution and artificial systems, human cognition and even
distribution of prime numbers (see \cite{hooge81,musha76,press78,koba82} and
references herein).

1/f noise is an intermediate between the well understood white noise with no
correlation in time and the random walk (Brownian motion) noise with no
correlation between increments. Widespread occurrence of signals exhibiting
power spectral density with 1/f behavior suggests that a general
mathematical explanation of such an effect might exist. However, except some
formal mathematical speculations like ''fractional Brownian motion'' or
half-integral of a white noise signal \cite{mandelbr68} no generally
recognized mathematical explanation of the ubiquity of 1/f noise is still
proposed. Physical models of 1/f noise in some physical systems are usually
very specialized, complicated (see \cite{hooge81,musha76,press78,koba82} and
references herein) and they do not explain the omnipresence of the processes
with $1/f^\delta $ spectrum \cite{icnf95,icnf97,upon97}.

Note also some mathematical algorithms and models of the generation of the
processes with 1/f noise \cite{jensen91,kumic94,sinha96}. These models also
expose some shortcomings: they are very specific, formal or unphysical, they
can not, as a rule, be solved analytically and they do not reveal the origin
as well as the necessary and sufficient conditions for the appearance of 1/f
type fluctuations.

History of the progress in different areas of physics indicates to the
crucial influence of simple models on the advancement of the understanding
of the main points of the new phenomena. We note here only the decisive
influence of the Bohr model of hydrogen atom on the development of the
quantum theory, the role of the Lorenz model as well as the logistic and
standard (Chirikov) maps for understanding of the deterministic chaos and
the quantum kicked rotator for the revealing the quantum localization of
classical chaos. On the contrary, the simple model of 1/f noise, to the best
of our knowledge, is absent until now. This is one of the reasons of the
incomprehensibility of the origin and nature of this phenomenon.

It is the purpose of this letter to present the simplest analytically
solvable model of 1/f noise which can essentially influence on the
understanding of the origin, main properties and parameter dependences of
the intensity and domination of the flicker noise. Our model is a result of
the search of necessary and sufficient conditions for the appearance of 1/f
fluctuations in simple systems affected by the random external perturbations
initiated in \cite{kaul97} and originated from the observation of the
transition from chaotic to nonchaotic behavior in the ensemble of randomly
driven systems \cite{kaul95}. Contrary to the McWhorter model \cite{surd39}
based on the superposition of large number of Lorentzian spectra and
requiring a very wide distribution of relaxation times, our model contains
only one relaxation rate and can have an exact 1/f spectrum in any desirable
wide range of frequency.

The simplest version of our model consists of one particle moving in the
closed contour. The period of the drift of the particle round the contour
fluctuates (due to the external random perturbations of the system's
parameters) about some average value $\tau $. So, a sequence of the transit
times $t_k$ when the particle crosses some point of the contour $L_m$ is
described by the recurrent equations
$$
\left\{
\begin{array}{ll}
t_k= & t_{k-1}+\tau _k, \\
\tau _k= & \tau _{k-1}-\gamma \left( \tau _{k-1}-\tau \right) +\sigma
\varepsilon _k.
\end{array}
\right. \eqno{(1)}
$$
Here $\gamma \ll 1$ is the period's relaxation rate, $\left\{ \varepsilon
_k\right\} $ denotes the sequence of uncorrelated normally distributed
random variables with zero expectation and unit variance (the white noise
source) and $\sigma $ is the standard deviation of white noise.

The intensity of the current of the particle through the section of the
contour $L_m$ is expressed as
$$
I\left( t\right) =\sum_k\delta \left( t-t_k\right) \eqno{(2)}
$$
where $\delta \left( t\right) $ is the Dirac delta function. The power
spectral density of the current (2) is
$$
S\left( f\right) =\lim \limits_{T\rightarrow \infty }\left\langle \frac 2T
\left| \sum_{k=k_{\min }}^{k_{\max }}e^{-i2\pi ft_k}\right| ^2\right\rangle
=\lim \limits_{T\rightarrow \infty }\left\langle \frac 2T\sum_k\sum_{q=k_{
\min }-k}^{k_{\max }-k}e^{i2\pi f\left( t_{k+q}-t_k\right) }\right\rangle
\eqno{(3)}
$$
where $T$ is the whole observation time interval, $k_{\min }$ and $k_{\max }$
are minimal and maximal values of index $k$ in the interval of observation
and the brackets $\left\langle ...\right\rangle $ denote the averaging over
realizations of the process.

From Eqs. (1) it follows an expression for the period
$$
\tau _k=\tau +\left( \tau _0-\tau \right) \left( 1-\gamma \right) ^k+\sigma
\sum_{j=1}^k\left( 1-\gamma \right) ^{k-j}\varepsilon _j\eqno{(4)}
$$
where $\tau _0$ is the initial period.

After some algebra we also can easily obtain an explicit expression for the
transit times $t_k$,
$$
t_k=t_0^{^{\prime }}+k\tau +\frac \sigma \gamma \sum_{l=1}^k\left[ 1-\left(
1-\gamma \right) ^{k+1-l}\right] \varepsilon _l.\eqno{(5)}
$$
Here $t_0^{^{\prime }}$ is some constant for $k\gg \gamma ^{-1}$ or $\tau
_0=\tau $. In the later case $t_0^{^{\prime }}$ is the initial time $t_0$.
At $k\gg \gamma ^{-1}$ Eq. (5) generates the stationary time series and the
transit times $t_{k+q}$ and $t_k$ difference in Eq. (3) is
$$
t_{k+q}-t_k=q\tau +\frac \sigma \gamma \left\{ \left[ 1-\left( 1-\gamma
\right) ^q\right] \sum_{l=1}^k\left( 1-\gamma \right) ^{k+1-l}\varepsilon
_l+\sum_{l=k+1}^{k+q}\left[ 1-\left( 1-\gamma \right) ^{k+q+1-l}\right]
\varepsilon _l\right\} ,q\geq 0.\eqno{(6)}
$$
Substitution of Eq. (6) into Eq. (3) and averaging over realizations of the
process or over the normal distribution of the random variables $\varepsilon
_l$ yield
$$
S\left( f\right) =\lim \limits_{T\rightarrow \infty }\frac 2T
\sum_k\sum_qe^{i2\pi f\tau q-\pi ^2\sigma ^2f^2g\left( q\right) }\eqno{(7)}
$$
where
$$
g\left( q\right) =\frac 2{\gamma ^2}\left\{ \left[ 1-\left( 1-\gamma \right)
^q\right] ^2\sum_{l=1}^k\left( 1-\gamma \right) ^{2l}+\sum_{l=1}^q\left[
1-\left( 1-\gamma \right) ^l\right] ^2\right\} ,\quad q\geq 0.\eqno{(8)}
$$
Summations in Eq. (8) for $k\gg \gamma ^{-1}$ result in
$$
g\left( q\right) =\frac 2{\gamma ^2}\left\{ q-2\frac{\left( 1-\gamma \right)
\left[ 1-\left( 1-\gamma \right) ^q\right] }{1-\left( 1-\gamma \right) ^2}
\right\} .\eqno{(9)}
$$
Expansion of expression (9) in powers of $\gamma q\ll 1$ is
$$
g\left( q\right) =\frac 1\gamma q^2-\frac 13q^3+\frac 12q^2+\ldots .\quad
q\geq 0.\eqno{(10)}
$$
Note the parity of the function $g(q)$, $g(-q)=g(q)$, following from Eqs.
(6)--(8) at $k-\left| q\right| \gg \gamma ^{-1}$.

For $f\ll f_\tau =\left( 2\pi \tau \right) ^{-1}$ and $f<f_2=2\sqrt{\gamma }
/\pi \sigma $ we can replace the summation in Eq. (7) by the integration
$$
S\left( f\right) =2\bar I\int\limits_{-\infty }^{+\infty }e^{i2\pi \tau
fq-\pi ^2\sigma ^2f^2g\left( q\right) }dq.\eqno{(11)}
$$
Here $\bar I=\lim \limits_{T\rightarrow \infty }\left( k_{\max }-k_{\min
}+1\right) /T=\tau ^{-1}$ is the averaged current. Furthermore, at $f\gg
f_1=\gamma ^{3/2}/\pi \sigma $ it is sufficient to take into account only
the first term of expansion (10). Integration in Eq. (11) hence yields to
1/f spectrum
$$
S\left( f\right) =\bar I^2\frac{\alpha _H}f,\quad f_1<f<f_2,f_\tau
\eqno{(12)}
$$
where $\alpha _H$ is a dimensionless constant (the Hooge parameter)
$$
\alpha _H=\frac 2{\sqrt{\pi }}Ke^{-K^2},\quad K=\frac{\tau \sqrt{\gamma }}
\sigma .\eqno{(13)}
$$

The model containing only one relaxation time $\gamma ^{-1}$ for
sufficiently small parameter $\gamma $ can, therefore, produce an exact
1/f-like spectrum in any desirable wide range of frequency. Furthermore, due
to the contribution to the transit times $t_k$ of the large number of the
random variables $\varepsilon _l$ ($l=1,2,...k$), our model represents a
'long-memory' random process. As a result of the nonzero relaxation rate ($
\gamma \neq 0$) and, consequently, due to the finite dispersion of the $\tau
$ period, $\sigma _\tau ^2\equiv \left\langle \tau _k^2\right\rangle
-\left\langle \tau _k\right\rangle ^2=\sigma ^2/2\gamma \left( 1-\gamma
/2\right) $ ($2k\gamma \gg 1$), the model, however, is free from the
unphysical divergency of the spectrum at $f\rightarrow 0$. So, using an
expansion of expression (9) at $\gamma q\gg 1$, $g(q)=2q/\gamma ^2$, we
obtain from Eq. (11) the spectrum density $S(f)=\bar I^2\left( 2\sigma
^2/\tau \gamma ^2\right) $ for $f\ll f_1,f_0=\tau \gamma ^2/\pi \sigma ^2$.

Eqs. (11)--(13) describe quite well the power spectrum of the random process
(1). As an illustrative example in Fig. 1 the numerically calculated power
spectral density averaged over five realizations of the process (1) is
compared with the analytical calculations according to Eqs. (9)--(13). Note,
that for the used in calculations parameters $\tau $, $\sigma $, and $\gamma
$ according to Eqs. (11)--(13) we have $K=2$, $\alpha _H=0.04$, $f_1=5\times
10^{-4}$, $f_2=0.06$, $f_\tau =0.08$, $f_0=1\times 10^{-3}$, while spectrum
for $f_1<f<f_2$ according to Eq. (12) is $S(f)=0.01/f$ and for $f<<f_1,f_0$
tends to the constant $S(f<<f_1)=156$. These results are in excellent
agreement with the numerical analysis.

This simple, consistent and exactly solvable model can easily be generalized
in different directions: for large number of particles moving in similar
contours with the coherent (identical for all particles) or independent
(uncorrelated for different particles) fluctuations of the periods, for the
non-Gaussian and for the continuous perturbations of the systems' parameters
and for some spatially extended systems. So, when an ensemble of $N$
particles moves in the closed contours and the period of each particle
fluctuates independently (due to the perturbations by uncorrelated,
different for each $\upsilon $ particle, sequences of random variables $
\{\varepsilon _k^\upsilon \}$) the power spectral density of the collective
current $I$ of all particles can be calculated by the above method too and
is expressed as the Hooge formula \cite{hooge81}
$$
S\left( f\right) =\bar I^2\frac{\alpha _H}{Nf}.\eqno{(14)}
$$

Summarizing, simple analytically solvable model of 1/f noise revealing main
features and parameter dependences of the power spectral density of the
noise is proposed. The model and its generalizations may essentially
influence on the understanding of the origin, the main properties and
parameter dependences of the intensity and domination of the flicker noise.

The author acknowledges stimulating discussions with Prof. R. Katilius,
Prof. A. Matulionis, and Dr. A. Bastys as well as numerical simulation data
provided by Mr. T. Me\v skauskas. The research of this publication was made
possible in part by support of the Alexander von Humboldt Foundation and
Lithuanian State Science and Studies Foundation.

\newpage\

\begin{center}
{\bf Captions to the figure of the paper }\vspace{1cm}

{\LARGE B. Kaulakys} \vspace{1cm}

{\LARGE ''Simple model of 1/f noise''}
\end{center}

\vspace{1cm}

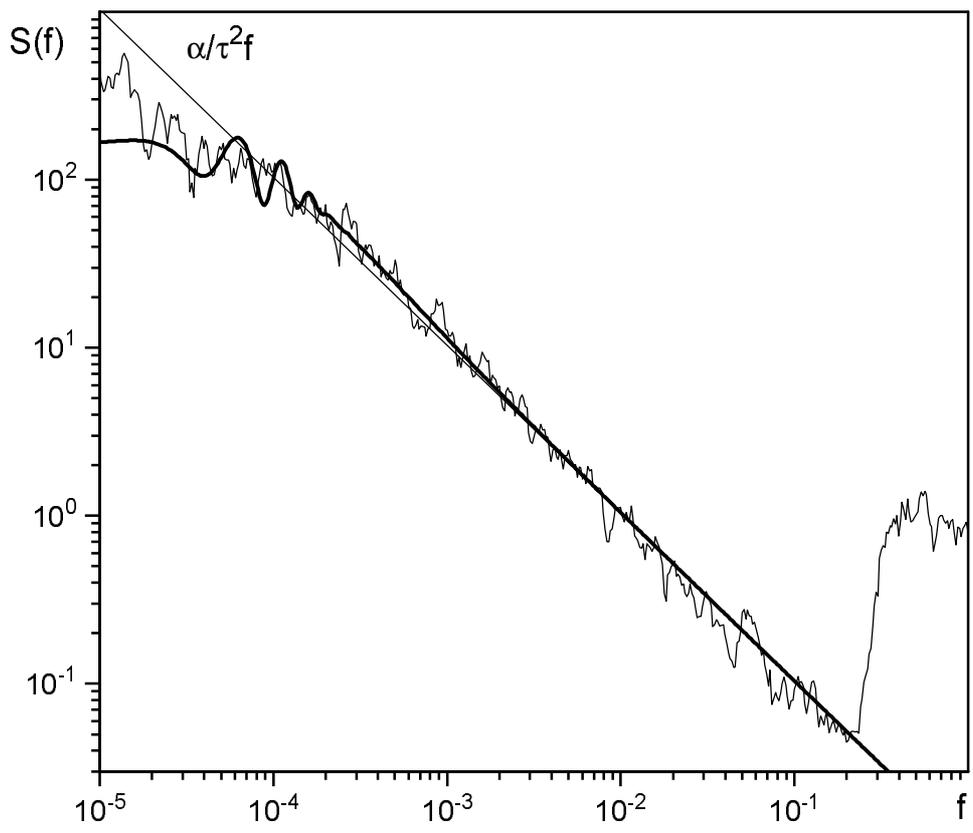
Fig.1. Power spectral density vs frequency of the current generated by Eqs.
(1) with $\tau =2$, $\sigma =0.04$, and $\gamma =0.016$. Sinuous fine curve
is the averaged over five realizations result of numerical simulations, the
heavy line corresponds to the numerical integration of Eq. (11) with $
g\left( q\right) $ from Eq. (9), and the thin straight line represents the
analytical spectrum according to Eqs. (12) and (13).

\end{document}